\pdfoutput=1

\documentclass[%
 reprint,
nofootinbib,
 amsmath,amssymb,
 aps,
]{revtex4-1}

\usepackage{graphicx}% Include figure files
\usepackage{dcolumn}% Align table columns on decimal point
\usepackage{bm}% bold math
\usepackage{xcolor}

\begin{document}

\preprint{APS/123-QED}

\title{TALEs from a spring -- superelasticity of Tal effector protein structures}

\author{Holger Flechsig}
\email{holgerflechsig@hiroshima-u.ac.jp}
\affiliation{Research Center for the Mathematics on Chromatin Live Dynamics (RcMcD)}
\affiliation{Department of Mathematical and Life Sciences, Graduate School of Science, Hiroshima University, 1-3-1 Kagamiyama, Higashi-Hiroshima, Hiroshima 739-8526, Japan}

\date{\today}

\begin{abstract}

 A simple force-probe setup is employed to study the mechanical properties of transcription activator-like effector (TALE) proteins in computer experiments.
It is shown that their spring-like arrangement benefits superelastic behaviour which is manifested by large-scale global conformational changes along the
helical axis, thus linking structure and dynamics in TALE proteins. As evidenced from the measured force-extension curves the dHax3 and PthXo1 TALEs
behave like linear springs, obeying Hooke's law, for moderate global structural changes. For larger deformations, however, the proteins exhibit nonlinearities
and the structures become stiffer the more they are stretched. Flexibility is not homogeneously distributed over TALE structure, but instead soft spots which
correspond to the RVD loop residues and present key agents in the transmission of conformational motions are identified.

\end{abstract}

\maketitle

\section*{Introduction}

TAL (transcription activator-like) effectors are proteins that are secreted in plants by bacteria of the {\it Xanthomonas} genus. Upon injection into
cells they are able to activate transcription of specific target plant genes which may be beneficial for bacterial infection \cite{kay_07,boch_10}.
Hence, a considerable amount of scientific attraction to this protein is owing to its role in the disease of various plant types including important
crops as well \cite{bogdanove_10,dodds_10}. Furthermore, artificial TALEs engineered to target prescribed sequences offer interesting applications
in genome editing \cite{scholze_11,boch_11,bogdanove_11}.

The common molecular architecture of TALEs consists of canonical two-helix repeats, each of them involved in the recognition of one specific DNA
base pair, that are arranged around a central axis to form an overall superhelical protein structure which wraps around a central groove which can
be occupied by duplex DNA \cite{deng_12,mak_12}.

Apparently, adequate understanding of the mechanisms underlying the operation in these important proteins requires a combination
of various experimental approaches, including structural determination and biochemical manipulation and analyses, with detailed methods
of molecular modelling. On the other side, the structure-function relationship of proteins, i.e. the principle of how the three dimensional folded
protein conformation defines its functional activity, may reveal surprisingly simple patterns. A prime example of such a case are molecular
machines and motors whose modular architecture, consisting of rigid domains connected by more flexible joints, gives rise to
well-organised relative internal motions through which the particular function is implemented \cite{alberts_98,vale_00}.

In the light of this perception one may ask the seemingly simple question which is probably most appealing from the perspective of a physicist:
Given the fact that TALEs look like a spring do they also exhibit spring-like dynamics and what are the benefits of such structure in terms of its
elastic properties? Those questions, which refer to the mechanical aspects of the operation of TALE proteins are addressed in this paper.

As we remember from high school physics we can probe the properties of deformable objects by holding it at one end and put a
weight on the other end. The force generated by the weight would then induce an extension of that object, i.e. a deviation
from its natural length, which can be easily measured. Using a variety of different weights allows to trace the dependence of
the extension from the applied force, a relation which is typically used to discuss the objects' elastic properties.

\begin{figure}[t!]
\centering{\includegraphics[width=7.5cm]{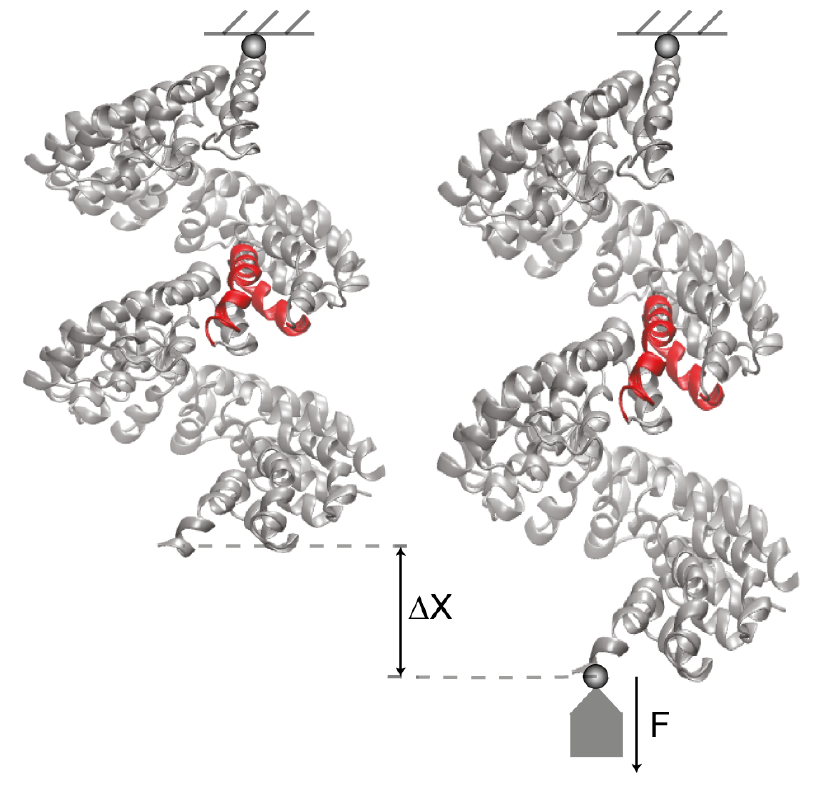}}
\caption{Schematic illustration of our {\it in silico} force-probe experiments. Exemplarily, the PthXo1 TAL effector is shown in
ribbon representation with one end being immobilised and the other end exerted to a force caused by a fictitious weight applied.
One selected TAL repeat is shown in red colour.}
\end{figure}

Here, we have performed such experiments in modelling simulations based on the currently available crystal structures of TALEs.
For the protein dynamics we have employed the coarse-grained elastic-network description, in which the protein is represented
as a network of beads connected via deformable strings \cite{tirion_96,bahar_97,haliloglu_97}. Despite gross simplifications present
in such models, they have been proven to perform remarkably well in the prediction of functional chemo-mechanical motions in proteins
\cite{tama_01,zheng_03,togashi_07,flechsig_10,flechsig_11}. It should be stressed that proteins modelled as elastic networks
still present highly complex systems which generally can only be treated numerically and, as shown previously, may exhibit strong
nonlinearities in their conformational dynamics \cite{togashi_10}.

Our analysis was performed on four different TALE structures. We have considered the DNA-complexed and the free form of the
artificially engineered dHax$3$ TALE \cite{deng_12}. Both structures contain $11$ TAL repeats which complete one helical turn,
but the corresponding pitch is found to be substantially different. Furthermore we have taken into account the structure of the PthXo1
TALE from the rice pathogen {\it Xanthomonas oryzae}, which was determined in the presence of duplex DNA and reveals an overall
two-turn helical shape formed by $23$ repeats \cite{mak_12}. To allow comparison with dHax$3$ we have also constructed a shortened
one-turn version of PthXo1 (named PthXo1* throughout the paper, see Methods section).

\section*{Methods}

We considered the structure of two TALE proteins, that of the artificially engineered dHax$3$ in its DNA-free form (PDB ID 3V6P)
and complexed with DNA (3V6T), and that of PthXo1 from the rice pathogen {\it Xanthomonas oryzae} bound to a DNA duplex (3UGM).
The elastic network was obtained by replacing each amino acid of the corresponding structure by a single bead that was placed at the
position of the respective alpha-carbon atom (denoted by $\vec{R}_{i}^{(0)}$ for bead $i$). Then each two beads were connected by a
deformable string if their spatial distance was below a prescribed interaction radius $r_{int}$. The total elastic energy of the network
$U=\sum_{i<j}^{N}\frac{A_{ij}}{2}(d_{ij}-d_{ij}^{(0)})^{2}$ is the sum over all string contributions, where $N$ is
the number of network beads and $A_{ij}=1$, if beads $i$ and $j$ are connected by a string, and $A_{ij}=0$ else. Here,
$d_{ij}^{(0)}=|\vec{R}_{i}^{(0)}-\vec{R}_{j}^{(0)}|$ is the natural length of a string connecting beads $i$ and $j$ (as extracted from the respective PDB file) and $d_{ij}=|\vec{R}_{i}-\vec{R}_{j}|$
is its corresponding deformed length, with $\vec{R}_{i}=(x_{i},y_{i},z_{i})^{T}$ being the actual position vector of bead $i$. The energy given above
represents a rescaled energy in which the dependency from the stiffness constant (which was the same for all strings) was removed.
Neglecting thermal fluctuations and hydrodynamical interactions,
the dynamics of the network can be described by a set of Newton equations considered in the over-damped limit \cite{togashi_07}.
For bead $i$ the equation of motion is $\gamma\frac{d}{dt}\vec{R}_{i}=\sum_{j}^{N}\vec{F}_{ij}+\vec{f}_{i}^{ext}$, where
$\vec{F}_{ij}=-\frac{\partial}{\partial\vec{R}_{i}}U=-A_{ij}\frac{d_{ij}-d_{ij}^{(0)}}{d_{ij}}(\vec{R}_{i}-\vec{R}_{j})$ are the internal
forces generated by deformed strings which are connected to bead $i$ and $\vec{f}_{i}^{ext}$ is an external force which can be applied to that bead. These equations are
numerically integrated to obtain the position of all network beads at every time moment. Using a rescaled time we can remove the
the friction coefficient $\gamma$ (the same for all beads) on the left hand side of the equations of motion. Due to the energy
rescaling the forces have the dimension of lengths in our description. Note that the network dynamics is generally nonlinear since distances
are nonlinear functions of the spatial coordinates, i.e. $d_{ij}=\sqrt{(x_{i}-x_{j})^{2}+(y_{i}-y_{j})^{2}+(z_{i}-z_{j})^{2}}$.

The constructed network of dHax3 consisted of $N=373$ beads (corresponding to Gly$^{303}$-Gly$^{675}$) and the PthXo1 network had
$N=789$ beads (Gly$^{234}$-Asp$^{1032}$). In the PthXo1* network the shortened structure (Gly$^{234}$-Cys$^{623}$) was
considered and $N=388$. All networks were constructed using an interaction radius of $8$\AA.

To implement our force-probe experiment we have immobilised the network at the bead with index $0$ (on one side of the protein) and applied an
external force to a single bead with index $f$ (on the other protein side). The force had magnitude $F$ and the direction was chosen
to coincide with a vector that connects the two selected beads, i.e. $\vec{f}_{f}^{ext}=F\vec{u}$ with the unit vector $\vec{u}=(\vec{R}_{f}-\vec{R}_{0})/d_{f0}$.
In the simulations we have varied the force magnitude $F$ and, each time starting from the initial network, integrated the equations of motions until a steady
conformation of the network with the applied force was reached. In that state we have measured the protein extension as $\Delta X=d_{f0}-d_{f0}^{(0)}$.

In the simulations we wanted to stretch the TALEs along the `spring axis', i.e. along the superhelical axis
which coincides with the orientation of the DNA groove. Therefore, the immobilised network bead (index $0$) and the bead to
which the force is applied (index f) were carefully chosen. For the dHax3 TALE (in the DNA-free structure as well as in the bound
conformation) the immobilised bead corresponded to residue Gly$^{303}$ at the beginning of the 1b TALE repeat\footnote[1]{notations
from the original publication are used} and the force was applied to the bead which corresponded to residue Gly$^{675}$,
which is located at the end of TALE repeat 11.5a. For PthXo1 and PthXo1* the immobilised bead was Gly$^{234}$ from the long helix of repeat $-1$\footnotemark[1].
In the case of PthXo1* the forced bead was Lys$^{609}$ from the long helix of repeat $10$ and for the entire PthXo1 TALE the
forced bead corresponded to residue Lys$^{1016}$ from the long helix of repeat $22$.

To investigate the distribution of deformations in the final deformed TALE networks, we have assigned each bead $i$ the value
$\chi_{i}=(\sum_{j}^{N}A_{ij}\cdot|d_{ij}-d_{ij}^{(0)}|)/\sum_{j}^{N}A_{ij}$, which is the average absolute deformation of strings
connected to bead $i$.

Figures of protein conformations were prepared with the VMD software \cite{humphrey_96}.

\section*{Results}

A schematic representation of our force-probe setup and the molecular architecture of the PtXo1 TALE is shown in Fig. 1. In the Methods
section a detailed description of the performed computer experiments is provided.

The force-extension curves obtained from our simulations are displayed in Fig. 2. They show that through the application of the prescribed forces
all four TALE structures were able to undergo significant stretching, revealing that they exhibit an intrinsic flexibility along their superhelical
axis. In the Supplementary Movie large-scale global structural changes from a single simulation of force-induced stretching of PthXo1 are shown.
Three of the TALE proteins could be stretched to far more than twice their initial lengths.
The DNA-free dHax$3$ protein presents an exception; here a force of the same magnitude could induce relative length changes of `only' $52\%$.
We have checked whether the identified force-induced strong deformations were reversible and found that in all cases, after the release of the applied force,
the TALE protein returned to its particular initial structure. This finding indicates that the overall structural arrangement of TALEs must possess remarkable
elastic properties which may give rise to large-amplitude conformational motions along the superhelical axis.

\begin{figure}[t!]
\centering{\includegraphics[width=8.3cm]{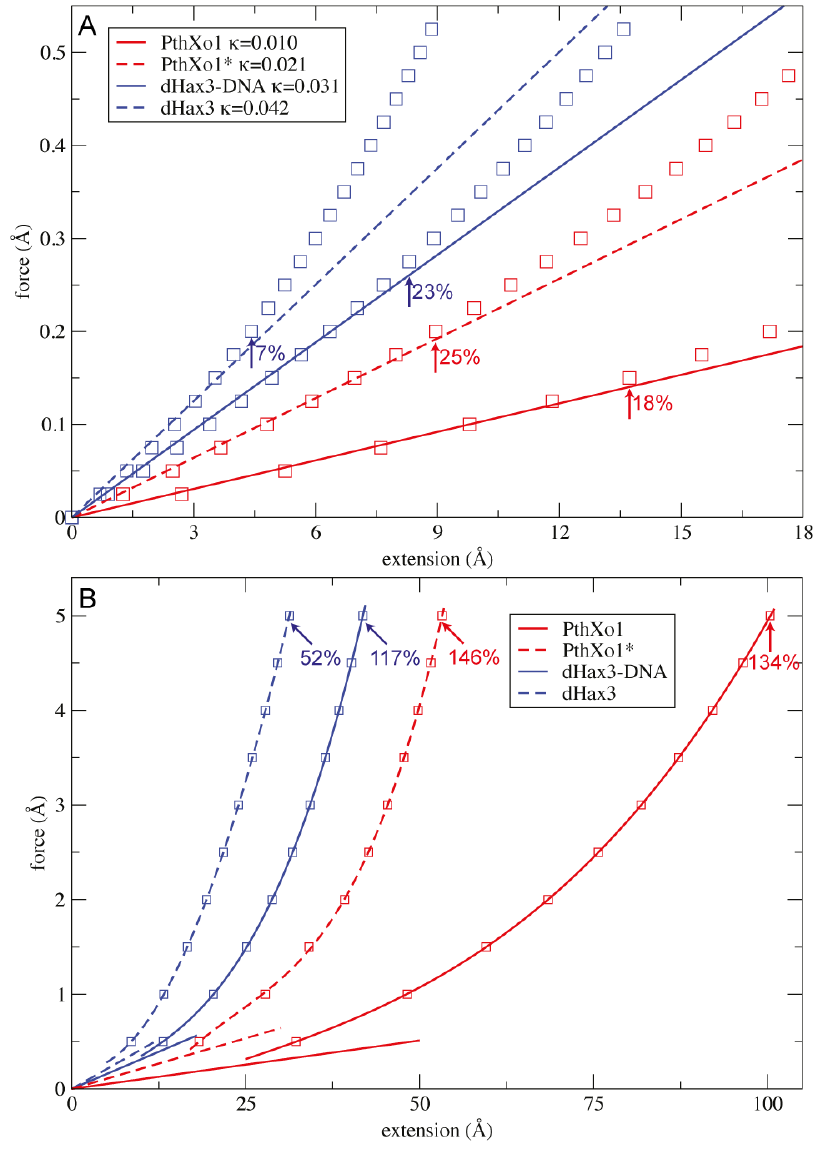}}
\caption{Force-extension relation for the four studied TALEs. A: Linear regression of the data and the derived stiffness constants are shown; the estimated validity range
in terms of relative extensions is indicated for each TAL structure. B: Non-linear super-proportional force extension curves for large TALE deformations. For the
force of magnitude $F=5.0$\AA, the relative protein extensions are given.}
\end{figure}

We observed that the dependence of the extension of TALE structures from the applied force is separated into two regimes.
For moderate extensions, i.e. when deviations from the initial length were not too large, a linear relation to the force is found, which represents
Hooke's law with the proportionality factor corresponding to the stiffness constant which can be assigned to the respective TALE structure. The validity of this linear behaviour differs for the four
TALEs (see Fig. 2); for the dHax$3$-DNA and PthXo1* the range (as estimated from the force-extension curves) is up to relative extensions of $25\%$.
Across that region and for deformations far beyond the
initial protein length we identified a nonlinear regime in which the force grows super-proportional with the extension, i.e. the structure gets stiffer the
more it is stretched.

The stiffness constants derived from a linear regression in the linear regime show that the full two-turn PthXo1 TALE is the softest structure as compared to the other TALEs.
The single-turn PthXo1* is twice as stiff as PthXo1 but softer than dHax$3$-DNA by a factor of $1.5$. The DNA-free dHax$3$ protein represents
the stiffest of the investigated TALE structures.

Apparently, the global large-scale elongations of TALE structures in response to forcing represent a collective effect resulting from the amplification of local
structural deformations which are still small. In the elongated states, as we find, such deformations are inhomogeneously spread over the TALE conformation and multiple
residues which define soft spots of the structure can be identified (results for dHax3 are shown in Fig. 3). They are located along the central DNA binding groove and belong
to the short RVD-loops which connect the two alpha-helices of each TAL repeat and are critical for establishing contacts to DNA. Similar observations are made
for PthXo1* and full PthXo1 (not shown in Fig.3 but in the Supplementary Movie the PthXo1 structure was coloured according to the deformation pattern of the elastic network).

\begin{figure}[t!]
\centering{\includegraphics[width=8.3cm]{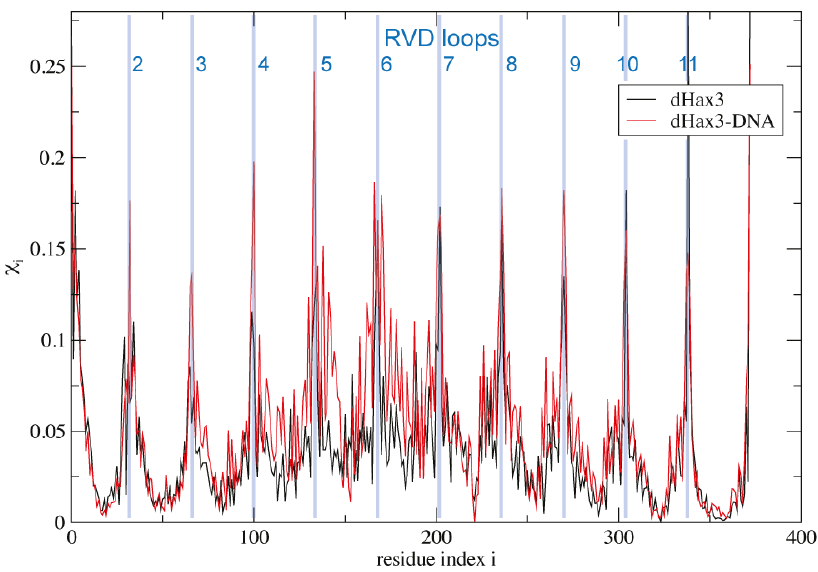}}
\caption{Distribution of local conformational changes in the elongated dHax3 structures obtained after application of an external force with magnitude $F=1.0$\AA.
The deformation value for the residues is plotted (see Methods, lines are used for better visibility). Positions of the RVD loops are indicated by blue lines.}
\end{figure}

We were curious whether conformational motions induced in our simple force-probe setup may cover aspects of the transition between the two
dHax$3$ structures, that with DNA and the DNA-free form. The DNA-bound structure has a length of $35.7$\AA\ (measured as the distance between
immobilised and forced residue, which roughly corresponds to the structural pitch, see Methods), whereas the same length of the DNA-free conformation
is $60.5$\AA. As we see from Fig. 2, the force that would be needed to generate the corresponding extension of $24.8$\AA\ in the dHax$3$-DNA structure cannot be
deduced from Hooke's law but instead necessitates the nonlinear relation. From a cubic regression of the data in that regime we computed the corresponding
magnitude of the force as $1.45$\AA. In a single simulation we have applied this force to the dHax$3$-DNA elastic network and compared the resulting extended
structure with that of the DNA-free dHax$3$ crystal structure (see Fig. 4). We find that after superposition of their C$\alpha$-atoms they compare with a
RMSD-value of $3$\AA.

\section*{Discussion}

In this paper we report results of very specific force-probe computer experiments performed for four Tal-effector (TALE) proteins. Their molecular
architecture shows a superhelical arrangements of basic structural repeats, and, crystal structures obtained for a particular TALE have indicated
pronounced conformational flexibility along the helical axis \cite{deng_12}. Inspired by their spring-like shape we aimed to investigate whether TALEs also exhibit
spring-like dynamical properties. For that purpose we have implemented a very basic setup of computer experiments in which a TALE protein was
immobilised at one end and exerted to a force at the opposite site. For the TALE dynamics we have employed the coarse-grained elastic-network
description in which the protein is viewed as a meshwork of beads connected via deformable strings. This approach emphasises the mechanical
aspects of protein function and allows for an efficient numerical implementation in computer experiments.

Our analysis, based on the evaluation of force-extension curves, shows that the investigated TALE structures indeed exhibit an enormous
flexibility along the superhelical axis and even very large deformations are found to be elastic, i.e. the structural changes are reversible.
The dynamical properties of TALEs resemble that of mechanical springs; in particular Hooke's law, i.e. the linear dependence of the extension
from the applied force is found to be valid for the considered protein structures. However, there are also differences to ordinary springs.
The elastic deformability of a macroscopic mechanical spring, typically designed as a periodically wound metal wire, results solely from its
particular helical shape and the spring material is usually stiff in itself. Proteins, however, represent {\it soft material} and the force-induced
global deformations in TALE structures must effectively result from a collective amplification phenomena manifested by the accumulation of local
conformational changes. To generate such internal motions requires forces acting on all protein residues and which each have to become
larger to induce an overall larger global deformation. While the linear regime is valid for moderate TALE extensions, the external force
applied in our simulations, which has to compensate all the internal forces, therefore grows in a super-proportional fashion for larger
deformations, meaning that the structures become stiffer the more they are stretched. Similar response curves are well-known for
mechanical progressive springs.

The stiffness constants for the four TALE structures obtained from the linear regime resemble what we know from macroscopic springs
\footnote[2]{It should be noted that due to the coarse-graining in the elastic network description the absolute scale of the dimensionless stiffness constants is arbitrary.}: A linear
spring that is cut in halves can be stretched by only half the fraction by the same force, i.e. it is twice as stiff. When comparing the full PthXo1 TALE, which
represents a long spring with two helical turns, with the artificially shortened PthXo1*, which has only one helical turn, we find this behaviour to be reproduced
also in the case of TALE protein springs. Comparing the DNA-bound PthXo1 and dHax3 TALEs is certainly complicated due to the lack of structural
data. Although our attempt revealed that PthXo1* has the softer structure, the underlying reasons can only be speculated about. Since both
structures comprise $11$ TAL repeats per helical turn and the pitch differs only by $2.4$\AA, the different elasticity may be owed to the fact
that dHax3 is an engineered protein whereas PthXo1 represents a structure evolved under real biological conditions. The dHax3 TALE
crystallised in the absence of DNA is found to be the stiffest spring and that for which the validity range of Hooke's law is most narrow.
This property can be explained by the fact that this structure, having a pitch of $60.5$\AA, already represents an extended spring as compared
to its DNA-bound version, which has a much lower pitch of $35.7$\AA. Hence, it responds to forcing by a larger stiffness.

Despite its approximate nature, our modelling could establish a link between the structural arrangement of TALE proteins and their dynamical properties
and demonstrate that the spring-like shape benefits superelasticity. In our description, amongst other simplifications, we neglected the effect of DNA in the
simulations (but we did so for both TALE proteins). It is therefore highly remarkable that our simple setup -- i.e. the TALE is hold at one end and a single
external force is applied at the opposite end -- can reproduce to a decent degree the global structural changes of the DNA-associated transition in the
dHax3 TALE. Furthermore our prediction that flexibility is not homogeneously distributed over the TALE structures, but instead soft regions are found
along the central DNA binding groove, is consistent with the crystallographic B-factors of the free dHax3 conformation, which are systematically
larger (i.e. residues are generally more mobile) for residues from the RVD loops. Our findings suggest that these soft spots represent key agents in the
transmission of conformational motions in TALE structures. The enhanced flexibility of RVD loops may be critical for the recognition and binding of bases
from the target DNA. Maybe such predictions can be checked in single molecule experiments. 

\begin{figure}[t!]
\centering{\includegraphics[width=8.3cm]{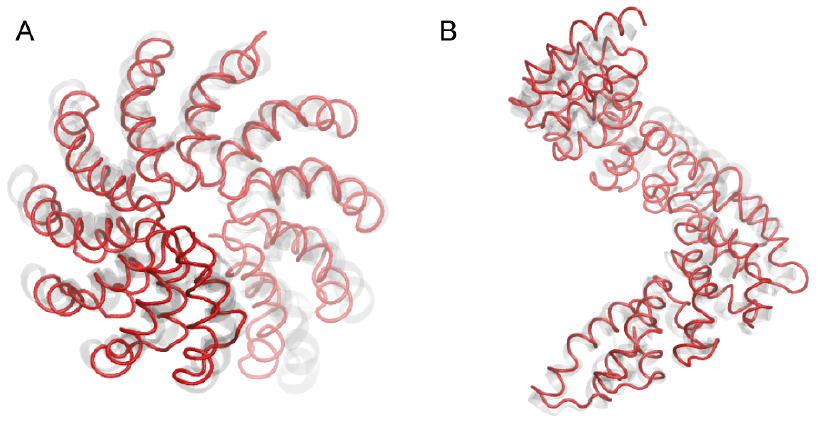}}
\caption{Comparison of dHax$3$ TALE structures. An {\it in silico} structure (shown as red tube) obtained from force-induced stretching of the DNA-bound
dHax$3$ conformation is superimposed with the DNA-free crystal structure of dHax$3$ (shown as transparent grey ribbons). Top view (A) and side view
perspective (B) are shown.}
\end{figure}

The super-elastic properties of TALE structures revealed by this study are apparently relevant for the functional activity of these proteins, since they
involve the ability to undergo enormous conformational changes along their superhelical axis. Indeed, the crystal structures of the artificially engineered
dHax3, representing the only TALE for which the DNA-bound and free protein could be determined to date, show that the free conformation is stretched by
$60\%$ as compared to the DNA-complexed form, thus providing evidence that the large-scale motions are linked to the interactions between the TALE and
DNA. Nuclear magnetic resonance studies together with small-angle X-ray scattering analysis performed for the effector protein PthA are in support of these
observations \cite{murakami_10} and a recently proposed model based on the analysis of AvrBs3 TALE mutants suggests relative motions of the TALE repeats
upon DNA target scanning and in the process of DNA recognition and binding to the RVDs \cite{schreiber_14}. Furthermore, detailed molecular dynamics simulation of the dHax3
TALE have shown that open-close motions between the two ends of the superhelical structure constitute the dominant conformational dynamics \cite{wan_13}.
Hence, from the current experimental and modelling studies the significance of conformational flexibility for binding DNA has emerged as a central aspect
of TALE function.
Certainly further investigations are needed to elucidate the operation principles of TAL effectors. Generally it should be possible to probe the flexibility and elastic
properties of TALE structures in experiments using e.g. appropriate atomic force microscopy setups.

\section*{Acknowledgments}

The author thanks Naoya Tochio and Yuichi Togashi for helpful discussions.


\begin{thebibliography}{}

\bibitem{kay_07} S. Kay, S. Hahn, E. Marois, G. Hause, U. Bonas (2007) A bacterial effector acts as a plant transcription factor and induces a cell size regulator.
Science 318:648-651.

\bibitem{boch_10} J. Boch, U. Bonas (2010) {\it Xanthomonas} AvrBs3 family-type III effectors: discovery and function.
Annu. Rev. Phytopathol. 48:419-436.

\bibitem{bogdanove_10} A.J. Bogdanove, S. Schornack, T. Lahaye (2010) TAL effectors: finding plant genes for disease and defense.
Curr. Opin. Plant Biol. 13:394-401.

\bibitem{dodds_10} P.N. Dodds, J.P. Rathjen (2010) Plant immunity: towards an integrated view of plant-pathogen interactions.
Nat. Rev. Genet. 11:539-548.

\bibitem{scholze_11} H. Scholze, J. Boch (2011) TAL effectors are remote controls for gene activation.
Curr. Opin. Microbiol. 14:47-53.

\bibitem{boch_11} J. Boch (2011) TALEs of genome targeting.
Nature Biotechnol. 29:135-136.

\bibitem{bogdanove_11} A.J. Bogdanove, D.F. Voytas (2011) TAL effectors: customizable proteins for DNA targeting.
Science 333:1843-1846.

\bibitem{deng_12} D. Deng, et al. (2012). Structural basis for sequence-specific recognition of DNA by TAL effectors.
Science 335:720-723.

\bibitem{mak_12} A. Nga-Sze Mak, P. Bradley, P.A. Cernadas, A.J. Bogdanove, B.L. Stoddard (2012). The crystal structure of
TAL effector PthXo1 bound to its DNA target.
Science 335:716-719.

\bibitem{alberts_98} B. Alberts (1998) The cell as a collection of protein machines: preparing the next generation of molecular biologists.
Cell 92:291-294.

\bibitem{vale_00} R.D. Vale, R.A. Milligan (2000) The way things move: looking under the hood of molecular motor proteins.
Science 288:88-95.

\bibitem{tirion_96} M.M. Tirion (1996) Large amplitude elastic motions in proteins from a single-parameter, atomic analysis.
Phys. Rev. Lett. 77:1905-1908

\bibitem{bahar_97} I. Bahar, A.R. Atilgan, B. Erman (1997) Direct evaluation of thermal fluctuations in proteins using a single-parameter
harmonic potential.
Fold. Des. 2:173-181

\bibitem{haliloglu_97} T. Haliloglu, I. Bahar, B. Erman (1997) Gaussian dynamics of folded proteins.
Phys. Rev. Lett. 79: 3090-3093

\bibitem{tama_01} F. Tama, Y. H. Sanejouand (2001) Conformational change of proteins arising from normal mode calculations.
Protein Engineering 14:1-6

\bibitem{zheng_03} W. Zheng, S. Doniach (2003) A comparative study of motor-protein motions by using a simple elastic-network model.
Proc. Natl. Acad. Sci. USA. 100:13253-13258

\bibitem{togashi_07} Y. Togashi, A.S. Mikhailov (2007) Nonlinear relaxation dynamics in elastic networks and design principles of
molecular machines.
Proc. Natl. Acad. Sci. USA. 104:8697-8702

\bibitem{flechsig_10} H. Flechsig, A.S. Mikhailov (2010) Tracing entire operation cycles of molecular motor hepatitis C virus helicase
in structurally resolved dynamical simulations.
Proc. Natl. Acad. Sci. USA. 107:20875-20880

\bibitem{flechsig_11} H. Flechsig, D. Popp, A.S. Mikhailov (2011) {\it In silico} investigation of conformational motions in superfamily 2 helicase
proteins.
PLoS ONE 6:e21809

\bibitem{togashi_10} Y. Togashi, T, Yanagida, A.S. Mikhailov (2010) Nonlinearity of mechanochemical motions in motor proteins.
PLoS Comput. Biol. 6:e1000814

\bibitem{humphrey_96} W. Humphrey, A. Dalke, K. Schulten (1996) VMD - Visual Molecular Dynamics.
J. Molec. Graphics 14:33-38

\bibitem{murakami_10} M.T. Murakami et al. (2010) The repeat domain of the type III effector protein PthA shows a TPR-like
structure and undergoes conformational changes upon DNA interaction.
Proteins 78:3386-3395

\bibitem{schreiber_14} T. Schreiber, U. Bonas (2014) Repeat $1$ of TAL effectors affects target specificity for the base at position zero.
Nucl. Acids Res. Advance Access May 3, 2014

\bibitem{wan_13} H. Wan, J-p. Hu, K-s. Li, X-h. Tian, S. Chang (2013) Molecular dynamics simulations of DNA-free and DNA-bound TAL effectors.
PLoS ONE 8:e76045

\end{thebibliography}
\end{document}